\documentclass[12pt]{article}  
\usepackage{amsmath,amssymb,amsfonts}
\usepackage{longtable}

\def\mylist{\begin{list}{}{\setlength{\leftmargin}{0.5in}
               \setlength{\listparindent}{-0.5in}
               \setlength{\itemindent}{\listparindent}}}

\newcommand{\ket}[1]{|#1\,\rangle}

\newcommand{\nD}[1]{\not D}
\newcommand{\cN}{{\mathcal N}}
\newcommand{\RR}{{\mathbb R}}
\newcommand{\CC}{{\mathbb C}}
\newcommand{\ZZ}{{\mathbb Z}}
\newcommand{\PP}{{\mathbb P}}

\newcommand{\Tr}{{\rm Tr}}

\newcommand{\eps}{\epsilon}

\newcommand{\da}{{\dot{a}}}
\newcommand{\db}{{\dot{b}}}
\newcommand{\sign}{{\rm sign}}

\newcommand{\D}{{\mathbf D}}

\begin{document}

\begin{titlepage}

\title{New and old $\cN=8$ superconformal field theories in three dimensions}
\author{Denis Bashkirov, Anton Kapustin\\ {\it California Institute of Technology}}

\maketitle

\abstract{We show that an infinite family of $\cN=6$ $d=3$ superconformal  Chern-Simons-matter theories has hidden $\cN=8$ superconformal symmetry and hidden parity on the quantum level. This family of theories is different from the one found by Aharony, Bergman, Jafferis and Maldacena, as well as from the theories constructed by Bagger and Lambert, and Gustavsson. We also test several conjectural dualities between BLG theories and ABJ theories by comparing superconformal indices of these theories.}

\end{titlepage}

\section{Introduction and summary}

Over the last few years several new classes of $\cN=8$ $d=3$ superconformal field theories have been discovered \cite{BL,G,ABJM}. Until then, it had been widely assumed that the only such theories are infrared limits of $\cN=8$ super-Yang-Mills theories and therefore are infinitely-strongly coupled. The newly discovered theories are not of this type. Rather they are Chern-Simons-matter theories which are superconformal already on the classical level. First of all, there are BLG theories \cite{BL,G} which have gauge group $SU(2)\times SU(2)$ \cite{vanR} and an arbitrary Chern-Simons coupling. $\cN=8$ supersymmetry in these theories is visible on the classical level. Then there are $\cN=8$  ABJM theories \cite{ABJM} which have gauge group $U(N)\times U(N)$ and have Chern-Simons coupling $k=1$ or $k=2$. These theories have $\cN=6$ supersymmetry on the classical level, and $\cN=8$ supersymmetry arises as a quantum effect. $\cN=8$ ABJM theories are strongly coupled, but they have a a weakly-coupled AdS-dual description in the large-N limit \cite{ABJM} and describe the physics of M2-branes.

In this paper we exhibit another class of $\cN=8$ $d=3$ superconformal Chern-Simons-matter theories. The theories themselves are not new: they are a special class of ABJ theories describing fractional M2-branes \cite{ABJ}. The gauge group of ABJ theories is $U(M)\times U(N)$ with Chern-Simons couplings $k$ and $-k$ for the two factors. These theories have $\cN=6$ superconformal symmetry on the classical level for all values of $M,N,$ and $k$. We will show that for $M=N+1$ and $k=\pm 2$ they have hidden $\cN=8$ supersymmetry on the quantum level. The same kind of arguments were used by us in \cite{BK} to show that ABJM theories with gauge group $U(N)_k\times U(N)_{-k}$ and $k=1,2$ have hidden $\cN=8$ supersymmetry. 

At first sight it might seem unlikely that ABJ theories may have $\cN=8$ supersymmetry for $N\neq M$. These theories are not parity-invariant on the classical level, while all hitherto known $\cN=8$ $d=3$ theories are parity-invariant. On the other hand, we know of no reason why $\cN=8$ supersymmetry should imply parity-invariance. We will see that $U(N+1)_2\times U(N)_{-2}$ theories do have hidden parity-invariance on the quantum level. The definition of the parity transformation involves a nontrivial duality on one of the gauge group  factors.

ABJ theories with $M=N+1$ and $k=2$ have the same moduli space as $U(N)_2\times U(N)_{-2}$ ABJM theories. Nevertheless we show that at least for $N=1$ and $N=2$ (and presumably for higher $N$) these two $\cN=8$ theories are not isomorphic. We do this by comparing superconformal indices \cite{BBMR} of both theories. The indices are computed using the localization method of \cite{Kim}. 

The existence of two non-isomorphic $\cN=8$ superconformal field theories with the moduli space $(\RR^8/\ZZ_2)^N/S_N$ is unsurprising from the point of view of M-theory. Such theories should describe $N$ M2-branes on an orbifold $\RR^8/\ZZ_2$, and it is well-known that there are exactly two such orbifolds differing by G-flux taking values in $H^4(\RR\PP^7,\ZZ)=\ZZ_2$ \cite{Sethi}.

The interpretation of Bagger-Lambert-Gustavsson  theories in terms of M2-branes is unclear in general. However, for low values of $k$ it has been proposed that BLG theories describe systems of two M2-branes on $\RR^8$ or $\RR^8/\ZZ_2$ \cite{LT,Distleretal,LP}. Such systems of M2-branes are also described by ABJM and ABJ theories \cite{ABJM,ABJ}. Thus we may reinterpret these proposals in field-theoretic terms as isomorphisms between certain BLG theories and ABJM or ABJ theories. We test these proposals by computing the superconformal indices of BLG theories and comparing them with those of ABJM and ABJ theories. Based on this comparison, we propose that the following $\cN=8$ theories are isomorphic on the quantum level:

\begin{itemize}
\item $U(2)_1\times U(2)_{-1}$ ABJM theory and $(SU(2)_1\times SU(2)_{-1})/\ZZ_2$ BLG theory
\item $U(2)_2\times U(2)_{-2}$ ABJM theory and $SU(2)_2\times SU(2)_{-2}$ BLG theory
\item $U(3)_2\times U(2)_{-2}$ ABJ theory and $(SU(2)_4\times SU(2)_{-4})/\ZZ_2$ BLG theory
\end{itemize}
The first two of these isomorphisms have been discussed in \cite{LP}. 

We provide further evidence for the first of these dualities by showing that on the quantum level $(SU(2)_1\times SU(2)_{-1})/\ZZ_2$ BLG theory has a free sector realized by monopole operators with minimal GNO charge. This sector has $\cN=8$ supersymmetry and can be thought of as a free $\cN=4$ hypermultiplet plus a free $\cN=4$ twisted hypermultiplet. Thus this BLG theory has not one but two copies of $\cN=8$ supersymmetry algebra, one acting on the free sector and one acting on the remainder. This quantum doubling of the $\cN=8$ supercurrent multiplet is required by duality, because $U(2)_1\times U(2)_{-1}$ theory also has such a doubling on the quantum level, as well as a free sector \cite{BK}. All these peculiar properties stem from the fact that the theory of $N$ M2-branes in flat space must have a free $\cN=8$ sector describing the center-of-mass motion. In the ``traditional'' approach to the theory of $N$ M2-branes via the  $U(N)$ $\cN=8$ super-Yang-Mills theory, this decomposition is apparent on the classical level (one can decompose all fields into trace and traceless parts which then do not interact, with the trace part being free). In the ABJM description of the same system this decomposition arises only on the quantum level \cite{BK}. For $N=2$ we also have a BLG description of the same system, and the existence of a free sector is again a quantum effect.

Superconformal index provides a simple tool for distinguishing $\cN=8$ theories which have the same moduli space. We can apply this method to other BLG theories which do not have an obvious interpretation in terms of M2-branes. For example, as noted in \cite{LP},  $SU(2)_k\times SU(2)_{-k}$ and $(SU(2)_{2k}\times SU(2)_{-2k})/\ZZ_2$ BLG theories have the same moduli space for all $k$ and one may wonder if they are in fact isomorphic. We compare the indices of these theories for $k=1,2$ and show that they are different. We also find that for $k=1$ both BLG theories have an extra copy of the $\cN=8$ supercurrent multiplet realized by monopole operators. This indicates that each of these theories decomposes as a product two $\cN=8$ SCFTs which do not interact with each other. For higher $k$ there is only one copy of the $\cN=8$ supercurrent multiplet.

This work was supported in part by the DOE grant DE-FG02-92ER40701.

\section{The moduli space}

Consider the family of $\cN=6$ Chern-Simons-matter theories constructed by Aharony, Bergman and Jafferis \cite{ABJ}. The gauge group of such a theory is $U(M)\times U(N)$, with Chern-Simons couplings $k$ and $-k$. If we regard it as an $\cN=2$ $d=3$ theory, then the matter consists of two chiral multiplets $A_a$, $a=1,2$ in the  representation $(M,{\bar N})$ and two chiral multiplets $B_\da,\da=1,2$ in the representation $({\bar M},N)$. The theory has a quartic superpotential 
$$
W=\frac{2\pi}{k}\eps^{ab}\eps^{\da\db} \Tr A_a B_\da A_b B_\db
$$
which preserves $SU(2)\times SU(2)$ symmetry as well as $U(1)_R$ R-symmetry. The chiral fields $A_a$ and $B_\da$ transform as $({\bf 2},{\bf 1})_1$ and $({\bf 1},{\bf 2})_1$ respectively. It was shown in \cite{ABJ} that the Lagrangian of such a theory has $Spin(6)$ symmetry which contains $Spin(4)=SU(2)\times SU(2)$ and $U(1)_R$ as subgroups. This implies that the action has $\cN=6$ superconformal symmetry, and the supercharges transform as a ${\bf 6}$ of $Spin(6)$ R-symmetry.

We wish to explore the possibility that on the quantum level some of these theories have $\cN=8$ supersymmetry. A necessary condition for this is that at a generic point in the moduli space of vacua the theory has $\cN=8$ supersymmetry. The moduli space can be parameterized by the expectation values of the fields $A_a$ and $B_\da$. Let us assume $M\geq N$ for definiteness. The superpotential is such that the expectation values can be brought to the diagonal form \cite{ABJ}:
$$
\langle {A_a}^i_j\rangle=a_{ja} \delta^i_j,\quad \langle {B_\da}^j_i\rangle=b_\da^j \delta^i_j\quad i=1,\ldots,M,\quad j=1,\ldots,N.
$$
Thus the classical moduli space is parameterized by $2N$ complex numbers $a_{ja}$ and $2N$ complex numbers $b^j_\da$ which together parameterize $\CC^{4N}$. Unbroken gauge symmetry includes a $U(M-N)$ factor which acts trivially on the moduli, as well as a discrete subgroup of $U(N)$. The low-energy effective action for the $U(M-N)$ gauge field is the Chern-Simons action at level $k'=k-\sign(k)(M-N)$. Thus along the moduli space the theory factorizes into a free theory describing the moduli and the topological $U(M-N)$ Chern-Simons theory at level $k'$.  Note that for $M-N>|k|$ the sign of $k'$ is different from that of $k$. This has been interpreted in \cite{ABJ} as a signal that for $M-N>|k|$ supersymmetry is spontaneously broken on the quantum level, and that the classical moduli space is lifted. Therefore from now on we will assume $M-N\leq |k|$. 

The putative $\cN=8$ supersymmetry algebra must act trivially on the topological sector, so we need to analyze for which $M,N,$ and $k$ the free theory of the moduli has $\cN=8$ supersymmetry. This theory is a supersymmetric sigma-model whose target space is the quotient of $\CC^N$ by the discrete subgroup of $U(N)$ which preserves the diagonal form of the matrices $A_a$ and $B_b$. This discrete subgroup is a semi-direct product of the permutation group $S_N$ and the $\ZZ_k^N$ subgroup of the maximal torus of $U(N)$ \cite{ABJ}. Thus the target space is $(\CC^4/\ZZ_k)^N/S_N$. The action of $\ZZ_k$ on $\CC^4$ is given by
$$
z_i\mapsto \eta z_i,\quad  i=1,\ldots,4,\quad \eta^k=1.
$$
Here $z_{1,2}$ are identified with $a_{ia}$, $a=1,2$, while $z_{3,4}$ are identified with $b^j_\da$, $\da=1,2$. 

Free $\cN=2$ sigma-model with target $\CC^4\simeq\RR^8$ has $\cN=8$ supersymmetry and $Spin(8)$ R-symmetry. Supercharges transform as ${\bf 8}_c$ of $Spin(8)$, while the moduli parameterizing $\RR^8$ transform as ${\bf 8}_v$. The above $\ZZ_k$ action on ${\bf 8}_v$ factors through the $Spin(8)$ action on the same space, and for $|k|
>2$ its commutant with $\ZZ_k$ is $U(4)$. $\ZZ_k$ itself can be identified with the $\ZZ_k$ subgroup of the $U(1)$ subgroup of $U(4)$ consisting of scalar matrices. Under the $U(4)$ subgroup ${\bf 8}_c$  decomposes as ${\bf 6}_0+{\bf 1}_2+{\bf 1}_{-2}$, and therefore for $|k|>2$ only ${\bf 6}_0$ is $\ZZ_k$-invariant. Thus for $|k|>2$ the moduli theory has only $\cN=6$ supersymmetry.

For $|k|=1,2$ the $\ZZ_k$ subgroup acts trivially on ${\bf 8}_c$, and therefore these two cases are the only ones for which the theory of moduli has $\cN=8$ supersymmetry. In view of the above, for $|k|=1$ we may assume that $M-N\leq 1$ while for $|k|=2$ we may assume $M-N\leq 2$.

For $N=M$ and $|k|=1,2$ it has been argued in \cite{ABJM} that the full theory has $\cN=8$ supersymmetry on the quantum level. The hidden symmetry currents are realized by monopole operators. This proposal has been proved using controlled deformation to weak coupling \cite{BK}; for other approaches see \cite{Gus,BKK,O}.

It remains to consider the case $0<M-N\leq |k|$ for $|k|=1,2$. Some of these theories are dual to the $\cN=8$ ABJM theories with $N=M$ and $k=1,2$. Indeed, it has been argued in \cite{ABJ} that for $M-N\leq |k|$ the theory with gauge group $U(M)_k\times U(N)_{-k}$ is dual to the theory with gauge group $U(2N-M+|k|)_{-k}\times U(N)_k$.  We will call it the ABJ duality.\footnote{Alternatively, the ABJ duality follows from the $\cN=3$ version of the Giveon-Kutasov duality applied to the $U(M)$ factor \cite{KWY3}. One can also verify that the $S^3$ partition functions of the dual ABJ theories agree \cite{KWY3}.} It maps $M-N$ to $|k|-(M-N)$ and $k$ to $-k$. Hence the ABJ theory with gauge group $U(N+1)_1\times U(N)_{-1}$ is dual to the ABJ theory with gauge group $U(N)_{-1}\times U(N)_1$. Similarly, the ABJ theory with gauge group $U(N+2)_2\times U(N)_{-2}$ is dual to the ABJ theory with gauge group $U(N)_{-2}\times U(N)_2$. 

The only remaining case is the ABJ theory with gauge group $U(N+1)_2\times U(N)_{- 2}$ and its parity-reversal. Each theory in this family is self-dual under the ABJ duality combined with parity. Put differently, the combination of naive parity and ABJ duality is a symmetry for all $N$, i.e. while these theories are not parity-invariant on the classical level, they have hidden parity on the quantum level. In the remainder of this paper we will argue that this family of theories in fact has hidden $\cN=8$ supersymmetry and is not isomorphic to any other known family of $\cN=8$ $d=3$ SCFTs. We will also present evidence that certain BLG theories with $k=1,2$ are isomorphic to $\cN=8$ ABJ and ABJM theories for $N=1,2$.

\section{Monopole operators and hidden $\cN=8$ supersymmetry}

In this section we will show that the ABJ theory with gauge group $U(N+1)_2\times U(N)_{-2}$ has hidden $\cN=8$ supersymmetry. We will follow the method of \cite{BK} to which the reader is referred for details. The main step is to demonstrate the presence of protected scalars with scaling dimension $\Delta=1$ which live in the representation ${\bf 10}_{-1}$ of the manifest symmetry group $Spin(6)\times U(1)_T$. Here $U(1)_T$ is the topological symmetry of the ABJ theory whose current 
$$
J^\mu=-\frac{k}{16\pi} \eps^{\mu\nu\rho}\left (\Tr F_{\nu\rho}+\Tr \tilde F_{\nu\rho}\right).
$$
is conserved off-shell. Once the existence of these scalars is established, acting on them with two manifest supercharges produces conserved currents with $\Delta=2$ transforming in the representation ${\bf 6}_{-1}$ of $Spin(6)\times U(1)_T$. Since conserved currents in any field theory form a Lie algebra, these currents together with their Hermitian-conjugate currents, $Spin(6)$ currents and the $U(1)_T$ current must combine into an adjoint of some Lie algebra containing $Spin(6)\times U(1)_T$ Lie algebra as a subalgebra. The unique possibility for such a Lie algebra is $Spin(8)$, which implies that the theory has $\cN=8$ supersymmetry. 

The existence of $\Delta=1$ scalars transforming in ${\bf 10}_{-1}$ is established using a controlled deformation of the theory compactified on $S^2$ to weak coupling. This deformation preserves $Spin(4)\times U(1)_R$ subgroup of $Spin(6)$ as well as $U(1)_T$. Decomposing ${\bf 10}_{-1}$ with respect to this subgroup, we find that it contains BPS scalars in $({\bf 3},{\bf 1})_{1,-1}$ of $Spin(4)\times U(1)_R\times U(1)_T$ and anti-BPS scalars in $({\bf 1},{\bf 3})_{-1,-1}$. Such BPS scalars cannot disappear as one changes the coupling (see appendix A for a detailed argument), so it is sufficient to demonstrate the presence of BPS scalars at extremely weak coupling. Note that the scaling dimension $\Delta$ of an operator is now reinterpreted as the energy of a state on $S^2$.

The BPS scalars we are looking for have nonzero $U(1)_T$ charge and therefore are monopole operators \cite{BKW}. At weak coupling monopole operators in ABJ theories are labeled by GNO ``charges'' $(m_1,\ldots,m_M)$ and $({\tilde m}_1,\ldots,{\tilde m}_N)$. GNO charges label spherically symmetric magnetic fields on $S^2$ and are defined up to the action of the Weyl group of $U(M)\times U(N)$ \cite{GNO}. They do not correspond to conserved currents and can be defined only at weak coupling. Their sum however is related to the $U(1)_T$ charge:
$$
Q_T=-\frac{k}{4}\left(\sum m_i+\sum \tilde m_i\right).
$$
Equations of motion of the ABJ theory imply that $\sum m_i=\sum \tilde m_i$, so $Q_T$ is integral for even $k$ but may be half-integral for odd $k$. We are interested in the case $Q_T=-1$, $k=2$, which implies 
$$
\sum m_i=\sum \tilde m_i=1.
$$

Consider a bare BPS monopole, i.e. the vacuum state, with GNO charges $m_1=\tilde m_1=1$ and all other GNO charges vanishing. This state has $\Delta=0$ but because of Chern-Simons terms it is not gauge-invariant (does not satisfy the Gauss law constraint). One can construct a gauge-invariant state by acting on the bare BPS monopole with two creation operators corresponding to the fields ${\bar A}^{1\tilde 1}_a$, $a=1,2$. These states are completely analogous to the BPS scalars for the $U(N)_2\times U(N)_{-2}$ ABJM theory constructed in \cite{BK} (see eq. (13) in that paper). The resulting multiplet of states transforms as $({\bf 3},{\bf 1})_{1,-1}$ of $Spin(4)\times U(1)_R\times U(1)_T$. It also has $\Delta=1$ and zero spin, since the creation operators for the field ${\bar A}$  with lowest energy have $\Delta=1/2$ and zero spin.

Similarly, by starting from an anti-BPS bare monopole with the same GNO charges and acting on it with two creation operators belonging to the fields $B_a^{1\tilde 1}$ we obtain anti-BPS scalars which transform in $({\bf 1},{\bf 3})_{-1,-1}$. One can also check that no other GNO charges give rise to BPS scalars with $\Delta=1$. In view of the above discussion this implies that the $U(N+1)_2\times U(N)_{-2}$ ABJ theory has hidden $\cN=8$ supersymmetry.

\section{Superconformal index and comparison with other $\cN=8$ theories}

One may question if $U(N+1)_2\times U(N)_{-2}$ ABJ theories are genuinely distinct from other known $\cN=8$ $d=3$ theories. The moduli space of such a theory is 
$(\CC^4/\ZZ_2)^N/S_N$, which is exactly the same as the moduli space of the $U(N)_2\times U(N)_{-2}$ ABJ theory. They differ in that along the moduli space the former theory has an extra topological sector described by $U(1)$ Chern-Simons theory at level $1$.  The latter theory is not quite trivial \cite{Wittensl2}, but it is very close to being trivial; for example, it does not admit any nontrivial local or loop observables. In any case, one could conjecture that even at the origin of the moduli space the two $\cN=8$ $d=3$ theories differ only by this decoupled topological sector. Some evidence in support of this conjecture is that BPS scalars in the two theories are in 1-1 correspondence, as we have seen in the previous section.

Fortunately, in the last few years there has been substantial progress in understanding superconformal $d=3$ gauge theories which allows us to compute many quantities exactly. One such quantity is the partition function on $S^3$ \cite{KWY1}; another one is the superconformal index on $S^2\times S^1$ \cite{BBMR,Kim}. The superconformal index receives contribution from BPS scalars as well as from other protected states with nonzero spin. In what follows we will compute the index for several low values of $N$ and verify that it is different for the two families of $\cN=8$ theories. The perturbative contribution to the superconformal index for ABJM theories has been computed in \cite{BM}; the contributions of sectors with a nontrivial GNO charge has been determined in \cite{Kim}. We will follow the approach of \cite{Kim}.

Bagger and Lambert \cite{BL} and Gustavsson \cite{G} constructed another infinite family of $\cN=8$ $d=3$ superconformal Chern-Simons-matter theories with gauge group $SU(2)\times SU(2)$  and matter in the bifundamental representation. More precisely, as emphasized in \cite{ABJM,LP}, there are two versions of BLG theories which have gauge groups $SU(2)_k \times SU(2)_{-k}$ or $(SU(2)_k\times SU(2)_{-k})/\ZZ_2$ where $k$ is an arbitrary natural number. The moduli space is $(\CC^4\times\CC^4)/\D_{2k}$ and $(\CC^4\times\CC^4)/\D_k$ respectively, where $\D_k$ is the dihedral group of order $2k$ \cite{LT,Distleretal,LP}. For large enough $k$ the moduli space is different from the moduli space of ABJ theories and so BLG theories cannot be isomorphic to any of them. However, for low values of $k$ there are some coincidences between moduli spaces which suggest that perhaps some of BLG theories are isomorphic to ABJ theories. 

One such case is $k=1$ and $G=(SU(2)\times SU(2))/\ZZ_2$. The moduli space is $(\CC^4\times\CC^4)/\ZZ_2$ where $\ZZ_2$ exchanges the two $\CC^4$ factors. It is natural to conjecture that this theory is isomorphic  to $U(2)_1\times U(2)_{-1}$ ABJM theory. A derivation of this equivalence was proposed in \cite{LP}. Another special case is $k=2$ and $G=SU(2)\times SU(2)$. In that case the moduli space is isomorphic to $(\CC^4/\ZZ_2\times\CC^4/\ZZ_2)/\ZZ_2$, where the first two $\ZZ_2$ factors reflect the coordinates on the two copies of $\CC^4$, while the third one exchanges them \cite{LT,Distleretal,LP}. This is the same moduli space as that of $U(2)_2\times U(2)_{-2}$ ABJM theory and $U(3)_2\times U(2)_{-2}$ ABJ theory. It was conjectured in \cite{LP} that this BLG theory is isomorphic to the $U(2)_2\times U(2)_{-2}$ ABJM theory. Finally, one can take $k=4$ and $G=(SU(2)\times SU(2))/\ZZ_2$. The moduli space is the same as in the previous case, so one could conjecture that this BLG theory is isomorphic to either the $U(2)_2\times U(2)_{-2}$ ABJM theory or the $U(3)_2\times U(2)_{-2}$ ABJ theory.

Below we will first of all compute the superconformal index for the $U(N)_2\times U(N)_{-2}$ ABJM theories and $U(N+1)_2\times U(N)_{-2}$ ABJ theories for $N=1,2$ and verify that although these theories have the same moduli space, they have different superconformal indices and therefore are not isomorphic. We will also compute the index for the special BLG theories with low values of $k$ discussed above and test the proposed dualities with the ABJM and ABJ theories. We will see that certain BLG theories have an additional copy of the $\cN=8$ supercurrent multiplet which is realized by monopole operators. In some cases this is predicted by dualities.

\subsection{$\cN=8$ ABJM vs. $\cN=8$ ABJ theories}

The superconformal index for a supersymmetric gauge theory on $S^2\times{\mathbb R}$ is defined as 

\begin{align}
{\cal I}(x, z_i)=Tr[(-1)^Fx^{E+j_3}\prod_{i}z_i^{F_i}]\label{ind}
\end{align}
where $F$ is the fermion number, $E$ is the energy, $j_3$ is the third component of spin and $F_i$ are flavor symmetry charges. The index receives contributions only from states satisfying $\{Q,Q^\dagger\}=E-r-j_3=0$, where $Q$ is one of the 32 supercharges and $r$ is a $U(1)$ $R$-charge. For details the reader is referred to \cite{BBMR,Kim}. 

The localization method \cite{Kim} enables one to express the index in a simple form\footnote{The formula is written for the case of zero anomalous dimensions of all fields which is true for all theories with at least $\cN=3$ supersymmetry.}

\begin{align}
{\cal I}(x, z_i)=\sum_{\{n_i\}}\int[da]_{\{n_i\}}x^{E_0(n_i)}e^{S_{CS}^0(n_i,a_i)}exp(\sum_{m=1}^\infty f(x^m, z_i^m, ma_i))
\end{align}
where the sum is over GNO charges, the integral whose measure depends on GNO charges is over a maximal torus of the gauge group, $E_0(n_i)$ is the energy of a bare monopole with GNO charges $\{n_i\}$, $S_{CS}^0(n_i)$ is effectively the weight of the bare monopole with respect to the gauge group and the function $f$ depends on the content of vector multiplets and hypermultiplets. For details see \cite{Kim}.

We computed the indices for the $U(2)_2\times U(1)_{-2}$ and $U(1)_2\times U(1)_{-2}$ theories up to the sixth order in $x$ and found the following pattern. In each topological sector the indices agree at the leading order in $x$ as a consequence of the identical spectra of BPS scalars of the lowest dimension. However, next-to-leading terms are different which signals nonequivalence of these theories. We summarize our results in tables 1 and 2 in Appendix B.

It is possible to single out contributions from different topological sectors by treating topological $U(1)_T$ symmetry as a flavor symmetry and introducing a new variable $z$ into the index. The result is a double expansion in $x$ and $z$ with powers of $z$ multiplying contributions of the appropriate topological charge . Alternatively, one can restrict summation over all GNO charges to those giving the desired topological charge. We used the second type of calculation.

We also compared the indices for the ABJ theory $U(3)_2\times U(2)_{-2}$ and the ABJM theory $U(2)_2\times U(2)_{-2}$ up to the fourth order in $x$. The contributions from different GNO sectors are summarized in tables 3 and 4 in Appendix B. Note that we count the contributions from the topological sectors $T\ge1$ twice because there is an identical contribution from the sectors with opposite topological charges. Starting at order $x^3$ the indices disagree, which means that these two $\cN=8$ theories, despite having the same moduli space, are not equivalent. 

\subsection{Comparison with BLG theories}

There are two BLG theories which have the same moduli space as $U(2)_2\times U(2)_{-2}$ ABJM and $U(3)_2\times U(2)_{-2}$ ABJ theories. They have gauge groups $SU(2)_2\times SU(2)_{-2}$ and $(SU(2)_4\times SU(2)_{-4})/\ZZ_2$. It is natural to conjecture that these four theories are pairwise isomorphic. Indeed, the moduli space is $(\CC^4/\ZZ_2\times\CC^4/\ZZ_2)/\ZZ_2$ in all four cases, suggesting that all these theories describe two M2-branes on an $\RR^8/\ZZ_2$ orbifold. It is well-known that there are two distinct $\RR^8/\ZZ_2$ orbifolds in M-theory \cite{Sethi}, which means that there should be only two nonisomorphic $\cN=8$ theories with this moduli space.

Comparison of the indices of the $U(2)_2\times U(2)_{-2}$ ABJM theory and the $SU(2)_2\times SU(2)_{-2}$ BLG theory (see Table 5) reveals their agreement up to the fourth order in $x$. Thus we conjecture that the two theories are equivalent. 

This conjecture can be checked further by comparing contributions to the indices from individual topological sectors on the ABJM side and sectors parametrized by the corresponding $U(1)$ charge on the BLG side. Recall that the topological charge $Q_T$ on the ABJM side is a charge of a $U(1)$ subgroup of the $Spin(8)$ $R$-symmetry group.  The commutant of this subgroup is $Spin(6)$ R-symmetry visible already on the classical level. Furthermore, the supercharge used in the deformation and the definition of the index is charged under a $U(1)$ subgroup of this $Spin(6)$.  On the BLG side, the whole $Spin(8)$ R-symmetry is visible on the classical level. Recall that  one can think of the BLG theory as a $\cN=2$ field theory with gauge group $SU(2)\times SU(2)$ and four chiral multiplets in the bifundamental representation. In this description, there is a manifest $SU(4)=Spin(6)$ symmetry under which the four chiral superfields transform as ${\bf 4}$. The commutant of this $Spin(6)$ symmetry is $U(1)_R$ symmetry with respect to which all four chiral superfields have charge $1/2$ and the supercharge has charge $1$. The topological charge $Q_T$ on the ABJM side corresponds to the charge of a $U(1)$ subgroup of $Spin(6)$ which we denote as $U(1)_t$\footnote{We now adopt the notation $T\equiv\sum_im_i$ for the topological charge and normalize the $U(1)_t$ charge of fundamental scalars of the BLG theories to $\pm1$ for notational convenience. The $U(1)_R$ charges are not shown in what follows.}. Thus we should compare the ABJM index in a particular topological sector with the BLG index in a sector with a particular $U(1)_t$ charge. The four chiral fields of the BLG theory decompose as ${\bf 4}={\bf 2}_{1}+{\bf 2'}_{-1}$ under $U(1)_t\times Spin(4)$.
%
To keep track of  $U(1)_t$ charges we introduce a new variable $z$ in accordance with (\ref{ind}). To the fourth order in $x$ only the $(\ket{0}\ket{0},\ket{1}\ket{1},\ket{2}\ket{2})$ GNO charges contribute. The two-variable index is
\begin{align}
& {\cal I}_{BLG,k=2}(x,z)=1+4x+21x^2+32x^3+53x^4+z^2(3x+16x^2+36x^3+48x^4)+\nonumber\\
& z^4(11x^2+36x^3+54x^4)+z^6(22x^3+64x^4)+45x^4z^8+ z^{-2}(3x+16x^2+36x^3+48x^4)+\nonumber\\ 
&z^{-4}(11x^2+36x^3+54x^4)+z^{-6}(22x^3+64x^4)+45x^4z^{-8}+O(x^5).
\end{align}
This is in a complete agreement with the index for the $U(2)_2\times U(2)_{-2}$ ABJM theory.

Similarly, we can compute the two-variable index for the $(SU(2)_4\times SU(2)_{-4})/\ZZ_2$ BLG theory. The difference compared to the $SU(2)\times SU(2)$ case is that the GNO charges are allowed to be half-integral, but their difference is required to be integral. The contributions of individual GNO charges are summarized in Table 6. We see that the total index agrees with that of the $U(3)_2\times U(2)_{-2}$ ABJ theory at least up to the fourth order in $x$. The two-variable index for this BLG theory is given by
\begin{align}
& {\cal I}'_{BLG}(x,z)=1+4x+21x^2+36x^3+39x^4+z^2(3x+16x^2+39x^3+40x^4)+\nonumber\\
& z^4(11x^2+36x^3+56x^4)+z^6(22x^3+64x^4)+45z^8x^4+z^{-2}(3x+16x^2+39x^3+40x^4)+\nonumber\\
& z^{-4}(11x^2+36x^3+56x^4)+z^{-6}(22x^3+64x^4)+45z^{-8}x^4+O(x^5)
\end{align}
and agrees with the two-variable index of the $U(3)_2\times U(2)_{-2}$ ABJ theory.

Lambert and Papageorgakis \cite{LP} argued that the $(SU(2)_1\times SU(2)_{-1})/\ZZ_2$ BLG theory is isomorphic to the $U(2)_1\times U(2)_{-1}$ ABJM theory. We can test this proposal in the same way by comparing the two-variable superconformal indices of the two theories. We find that they agree up to at least the fourth order in $x$. The contributions from different GNO charges are written down in Tables 7 and 8. They happen to match in each GNO sector separately. For a fixed topological charge on the ABJM side and the corresponding value of the $U(1)_t$ charge on the BLG side which manifests itself in the index as a power of $z$, the contribution to the index comes from a sum over different GNO charges, and the two sums happen to coincide term by term. For example, in the topological sector $T=1$ on the ABJM side the contribution from the GNO charge $\ket{n,1-n}\ket{n,1-n}$ equals the contribution from the GNO charge $\ket{n-1/2}\ket{n-1/2}$ with the first power of $z$ on the BLG side. 

The index makes apparent a peculiar feature of these two theories: they have twice the number of BPS scalars needed to enhance supersymmetry from ${\cal N}=6$ to ${\cal N}=8$. The first set of scalars has vanishing GNO charge. The corresponding contribution to the index is $\Delta{\cal I}=4x+3xz^2+3xz^{-2}$. It represents the decomposition ${\bf 10}={\bf 4}_0+{\bf 3}_2+{\bf 3}_{-2}$ under $U(1)_t\times Spin(4)\subset Spin(6)$.  The corresponding operators are gauge-invariant bilinear combinations of four chiral superfields present in the BLG model. The second set of ten BPS scalars comes from the GNO charge $\ket{1}\ket{1}$ and makes an identical contribution to the index. Ten BPS states are obtained by acting with ten scalar bilinears on the bare monopole to form gauge-invariant states $Q^{(i}Q^{j)}\ket{1}\ket{1}$. Here $Q^i$ is an off-diagonal component of the $i^{\rm th}$ complex scalar, $i=1,\ldots,4$. Among these ten states there are representations ${\bf (3,1)}_{1,-1}+{\bf (1,3)}_{1,1}$ of $Spin(4)\times U(1)_R\times U(1)_t$ with the normalization of the $U(1)_t$ charge as on page 4. Together with their Hermitian-conjugates, these BPS scalars lead to supersymmetry enhancement as in \cite{BK}.

The existence of two copies of the $\cN=8$ supersymmetry algebra for the $U(2)_1\times U(2)_{-1}$ ABJM theory was noted in \cite{BK}. It was shown there that the extra copy arises because the theory has a free sector with $\cN=8$ supersymmetry realized by monopole operators. The same is true about the $(SU(2)_1\times SU(2)_{-1})/\ZZ_2$ BLG theory, giving further support for the duality. The sector with the GNO charge $\ket{1/2}\ket{1/2}$ contains four gauge-invariant BPS scalars $Q^{i}\ket{1/2}\ket{1/2}$ with energy $\Delta=1/2$ whose contribution to the index is $\Delta{\cal I'}=2x^{1/2}z+2x^{-1/2}z$. This expression corresponds to the decomposition ${\bf 4}={\bf 2}_1+{\bf 2'}_{-1}$ under $U(1)_t\times Spin(4)\subset Spin(6)$. By virtue of state-operator correspondence these states correspond to four free fields with conformal dimension $\Delta=1/2$. Their bilinear combinations give rise to ten BPS scalars with GNO charge  $\ket{1}\ket{1}$ discussed above. This is in a complete agreement with the structure of the $U(2)_1\times U(2)_{-1}$ ABJM theory explored in \cite{BK}.

We can also use superconformal index to test whether certain BLG theories with identical moduli spaces are isomorphic on the quantum level. It has been noted in \cite{LP} that the moduli spaces of $SU(2)_k\times SU(2)_{-k}$ and $(SU(2)_{2k}\times SU(2)_{-2k})/\ZZ_2$ BLG theories are the same (they are both given by $(\CC^4\times\CC^4)/\D_{2k}$. We have seen above that for $k=2$ these two theories are not isomorphic. We also computed the index for $k=1$ and found that the indices disagree already at the second order in $x$ (Tables 9 and 10), so the theories are not equivalent. Examining BPS scalars, we find that neither of these theories has a free sector, but they both have two copies of the $\cN=8$ supercurrent multiplet.  One copy is visible on the classical level, while the BPS scalars of the other copy carry GNO charges, so it is  intrinsically quantum-mechanical in origin. The presence of the second copy of $\cN=8$ superalgebra indicates that on the quantum level both of these theories decompose into two $\cN=8$ SCFTs which do not interact with each other. This phenomenon does not occur for higher $k$.



\section*{Appendix A}

Our method of detecting hidden supersymmetry is based on deforming the theory to weak coupling and analyzing the spectrum of BPS scalars. 
In this appendix we provide a sufficient condition for BPS scalars to be protected as one deforms the coupling from weak to strong. In general, a local operator (or the corresponding state in the radial quantization) which lives in a short representation of the superconformal algebra can pair up with another short multiplet to form a long multiplet; quantum number of a long multiplet can change continuously as one deforms the coupling. We would like to show that this cannot happen for the cases of interest to us.

The kind of short multiplet we are interested in has a BPS scalar among its primaries. In the radial quantization such a state has energy $\Delta$ equal to its $U(1)_R$ charge $r$. To form a long multiplet there must be a short multiplet containing a spinor with energy $\Delta'=\Delta\pm 1/2$ and $R$-charge $r=r'\pm 1$. The option with $\Delta'=\Delta+1/2$ and $r'=r+1$ is ruled out by unitarity constraints \cite{BBMR}. These constraints also specify the short multiplet with the spinor. This is a so-called ``regular short multiplet" \cite{BBMR} with a scalar $\Delta''=\Delta-1$, $r''=r-2$ as the superconformal primary state satisfying $\Delta''=r''+1$. The zero-norm state is also a scalar, appears on the second level and has the quantum numbers of a BPS scalar $\Delta=r$. The spinor itself is on the first level. 

We conclude that a necessary condition for a BPS scalar with quantum numbers $\Delta=r$ to pair up into a long multiplet and flow away is the existence of a "regular short multiplet" with quantum numbers $\Delta''=\Delta-1$ and $r''=r-2$.

In the particular case of a $U(N+1)_k\times U(N)_{-k}$ ABJ theory and $\Delta=1$ such ``regular short multiplets" do not exist because $\Delta''=0$ and all physical states have $\Delta\geq 1$.

\newpage

\section*{Appendix B: Superconformal indices for $\cN=8$ ABJM, ABJ and BLG theories}

\begin{longtable}{|l|l|}
 \hline
GNO charges & Index contribution\\
  \hline 
 $T=0$ & $1+4x+2x^2+15x^4-16x^5+11x^6$\\
 \hline
 $\ket{0,0}\ket{0}$ & $1+4x+2x^2+15x^4-16x^5+2x^6$\\
 $\ket{1,-1}\ket{0}$ & $9x^6$\\
   \hline 
  $T=1$ & $3x+x^2-4x^3+20x^4-32x^5+24x^6$\\ 
  \hline
$\ket{1,0}\ket{1}$ & $3x+x^2-4x^3+20x^4-32x^5+24x^6$\\
\hline
$T=2$ & $5x^2+4x^3-5x^4+4x^5-4x^6$\\
\hline
$\ket{2,0}\ket{2}$ & $5x^2+4x^3-5x^4+4x^5-4x^6$\\
\hline
$T=3$ & $7x^3+4x^4+x^6$\\
\hline
$\ket{3,0}\ket{3}$ & $7x^3+4x^4+x^6$\\
\hline
$T=4$ & $9x^4+4x^5$\\
\hline
$\ket{4,0}\ket{4}$ & $9x^4+4x^5$\\
\hline
$T=5$ & $11x^5+4x^6$\\
\hline
$\ket{5,0}\ket{5}$ & $11x^5+4x^6$\\
\hline
$T=6$ & $13x^6$\\
\hline
$\ket{6,0}\ket{6}$ & $13x^6$\\
\hline
total & $1+10x+14x^2+14x^3+71x^4-42x^5+39x^6$\\
\hline
\caption{$U(2)_2\times U(1)_{-2}$. $T$ stands for the topological charge.}
\end{longtable}

\begin{longtable}{|l|l|}
\hline
Topological charge & Index contribution\\
\hline 
$T=0$ & $1+4x+x^2+4x^3+7x^4-12x^5+26x^6$\\
\hline   
$T=1$& $3x+4x^2+8x^4-4x^5+8x^6$\\ 
\hline
$T=2$& $5x^2+4x^3=8x^5-4x^6$\\
\hline
$T=3$& $7x^3+4x^4+8x^6$\\
\hline
$T=4$ & $9x^4+4x^5$\\
\hline
$T=5$ & $11x^5+4x^6$\\
\hline
$T=6$ & $13x^6$\\
\hline
total & $1+10x+19x^2+26x^3+49x^4+26x^5+92x^6$\\
\hline
\caption{$U(1)_2\times U(1)_{-2}$}
\end{longtable}

\begin{longtable}{|l|l|}
  \hline
GNO charges & Index contribution\\
  \hline 
   $T=0$ & $1+4x+21x^2+36x^3+39x^4$\\
 \hline
 $\ket{0,0,0}\ket{0,0}$ & $1+4x+12x^2+12x^3+5x^4$\\
 $\ket{1,0,-1}\ket{1,-1}$ & $9x^2+24x^3+10x^4$\\
 $\ket{2,0,-2}\ket{2,-2}$ & $25x^4$\\
 $\ket{1,0,-1}\ket{0,0}$ & $-x^4$\\    
  \hline 
  $T=1$ & $3x+16x^2+39x^3+40x^4$\\ 
  \hline
$\ket{1,0,0}\ket{1,0}$ & $3x+16x^2+24x^3+8x^4$\\
$\ket{2,0,-1}\ket{2,-1}$ & $15x^3+32x^4$\\
\hline
$T=2$ & $11x^2+36x^3+56x^4$\\
\hline
$\ket{1,1,0}\ket{1,1}$ & $6x^2+12x^3+9x^4$\\
$\ket{2,0,0}\ket{2,0}$ & $5x^2+24x^3+26x^4$\\
$\ket{3,0,-1}\ket{3,-1}$ & $21x^4$\\
\hline
$T=3$ & $22x^3+64x^4$\\
\hline
$\ket{2,1,0}\ket{2,1}$ & $15x^3+32x^4$\\
$\ket{3,0,0}\ket{3,0}$ & $7x^3+32x^2$\\
\hline
$T=4$ & $45x^4$\\
\hline
$\ket{2,2,0}\ket{2,2}$ & $15x^4$\\
$\ket{3,1,0}\ket{3,1}$ & $21x^4$\\
$\ket{4,0,0}\ket{4,0}$ & $9x^4$\\
\hline
total & $1+10x+75x^2+230x^3+445x^4$\\
\hline
\caption{$U(3)_2\times U(2)_{-2}$.
$T$ stands for the topological charge.}
\end{longtable}

\begin{longtable}{|l|l|}
  \hline
GNO charges & Index contribution\\
  \hline 
 $T=0$ & $1+4x+21x^2+32x^3+53x^4$ \\
 \hline
 $\ket{0,0}\ket{0,0}$ & $1+4x+12x^2+8x^3+12x^4$\\
 $\ket{1,-1}\ket{1,-1}$ & $9x^2+24x^3+16x^4$\\
 $\ket{2,-2}\ket{2,-2}$ & $25x^4$\\   
  \hline 
  $T=1$ & $3x+16x^2+36x^3+48x^4$\\ 
  \hline
$\ket{1,0}\ket{1,0}$ & $3x+16x^2+21x^3+16x^4$\\
$\ket{2,-1}\ket{2,-1}$ & $15x^3+32x^4$\\
\hline
$T=2$ & $11x^2+36x^3+54x^4$\\
\hline
$\ket{1,1}\ket{1,1}$ & $6x^2+12x^3+12x^4$\\
$\ket{2,0}\ket{2,0}$ & $5x^2+24x^3+21x^4$\\
$\ket{3,-1}\ket{3,-1}$ & $21x^4$\\
\hline
$T=3$ & $22x^3+64x^4$\\
\hline
$\ket{2,1}\ket{2,1}$ & $15x^3+32x^4$\\
$\ket{3,0}\ket{3,0}$ & $7x^3+32x^4$\\
\hline
$T=4$ & $45x^4$\\
\hline
$\ket{2,2}\ket{2,2}$ & $15x^4$\\
$\ket{3,1}\ket{3,1}$ & $21x^4$\\
$\ket{4,0}\ket{4,0}$ & $9x^4$\\
\hline
total & $1+10x+75x^2+220x^3+475x^4$\\
\hline
\caption{$U(2)_2\times U(2)_{-2}$.
$T$ stands for the topological charge.}
\end{longtable}

\begin{longtable}{|l|l|}
  \hline
GNO charges & Index contribution\\
  \hline 
$\ket{0}\ket{0}$ & $1+10x+40x^2+76x^3+114x^4$\\
$\ket{1}\ket{1}$ & $35x^2+144x^3+196x^4$\\ 
$\ket{2}\ket{2}$ & $165x^4$\\ 
\hline 
total & $1+10x+75x^2+220x^3+475x^4$\\ 
\hline
\caption{$SU(2)_2\times SU(2)_{-2}$}
\end{longtable}

\begin{longtable}{|l|l|}
  \hline
GNO charges & Index contribution\\
  \hline 
  $\ket{0}\ket{0}$ & $1+4x+12x^2+8x^3+12x^4+$\\
 & $z^2(3x+8x^2+12x^3+8x^4)+z^{-2}(3x+8x^2+12x^3+8x^4)+$\\
    & $z^4(6x^2+12x^3+12x^4)+z^{-4}(6x^2+12x^3+12x^4)+$\\
   & $z^6(10x^3+16x^4)+z^{-6}(10x^3+16x^4)+15z^8x^4+15z^{-8}x^4$\\
\hline
 $\ket{1/2}\ket{1/2}$ & $9x^2+28x^3+2x^4+$\\
  & $z^2(8x^2+27x^3+8x^4)+z^{-2}(8x^2+27x^3+8x^4)+$\\
  & $z^4(5x^2+24x^3+23x^4)+z^{-4}(5x^2+24x^3+23x^4)+$\\
  & $z^6(12x^3+32x^4)+z^{-6}(12x^3+32x^4)+21z^8x^4+21z^{-8}x^4$\\
\hline 
$\ket{1}\ket{1}$ & $25x^4+24z^2x^4+24z^{-2}x^4+24z^4x^4+24z^{-4}x^4+16z^6x^4+16z^{-6}x^4+9z^8x^4+9z^{-8}x^4$\\
 \hline
\caption{$(SU(2)_4\times SU(2)_{-4})/\ZZ_2$}
\end{longtable}

\begin{longtable}{|l|l|}
\hline
  GNO charges & Index contribution\\
  \hline 
  $\ket{0}\ket{0}$ & $1+4x+12x^2+8x^3+12x^4+$\\
 & $z^2(3x+8x^2+12x^3+8x^4)+z^{-2}(3x+8x^2+12x^3+8x^4)+$\\
    & $z^4(6x^2+12x^3+12x^4)+z^{-4}(6x^2+12x^3+12x^4)+$\\
   & $z^6(10x^3+16x^4)+z^{-6}(10x^3+16x^4)+15z^4x^4+15z^{-8}x^4$\\
\hline
 $\ket{1/2}\ket{1/2}$ & $2z(x^{\frac12}+6x^{\frac32}+10x^{\frac52}+7x^{\frac72})+2z^{-1}(x^{\frac12}+6x^{\frac32}+10x^{\frac52}+7x^{\frac72})+$\\  
  & $2z^3(3x^{\frac32}+10x^{\frac52}+9x^{\frac72})+2z^{-3}(3x^{\frac32}+10x^{\frac52}+9x^{\frac72})+$\\
  & $2z^5(6x^{\frac52}+14x^{\frac72})+2z^{-5}(6x^{\frac52}+14x^{\frac72})$\\
\hline 
 $\ket{1}\ket{0}$ & $-x^4$\\
\hline
$\ket{0}\ket{1}$ & $-x^4$\\
\hline
$\ket{1}\ket{1}$ & $4x+16x^2+16x^3+33x^4+$\\
 & $z^2(3x+16x^2+19x^3+24x^4)+z^{-2}(3x+16x^2+19x^3+24x^4)+$\\
 & $z^4(8x^2+24x^3+16x^4)+z^{-4}(8x^2+24x^3+16x^4)+$\\
 & $z^6(15x^3+32x^4)+z^{-6}(15x^3+32x^4)+24z^8x^4+24z^{-8}x^4$\\
 \hline
$\ket{3/2}\ket{3/2}$ & $2z(3x^{\frac32}+10x^{\frac52}+8x^{\frac72})+2z^{-1}(3x^{\frac32}+10x^{\frac52}+8x^{\frac72})+$\\
 & $2z^3(2x^{\frac32}+10x^{\frac52}+10x^{\frac72})+2z^{-3}(2x^{\frac32}+10x^{\frac52}+10x^{\frac72})+$\\
 & $2z^5(5x^{\frac52}+14x^{\frac72})+2z^{-5}(5x^{\frac52}+14x^{\frac72})+18z^7x^{\frac72}+18z^{-7}x^{\frac72}$\\
\hline
$\ket{2}\ket{2}$ & $9x^2+24x^3+16x^4+$\\
 & $z^2(8x^2+24x^3+16x^4)+z^{-2}(8x^2+24x^3+16x^4)+$\\
 & $z^4(5x^2+24x^3+21x^4)+z^{-4}(5x^2+24x^3+21x^4)+$\\
 & $z^6(12x^3+32x^4)+z^{-6}(12x^3+32x^4)+21z^8x^4+21z^{-8}x^4$\\
 \hline
 $\ket{5/2}\ket{5/2}$ & $2z(6x^{\frac52}+14x^{\frac72})+2z^{-1}(6x^{\frac52}+14x^{\frac72})+$\\
 & $2z^3(5x^{\frac52}+14x^{\frac72})+2z^{-3}(5x^{\frac52}+14x^{\frac72})+$\\
 & $2z^5(3x^{\frac52}+14x^{\frac72})+2z^{-5}(3x^{\frac52}+14x^{\frac72})+14z^7x^{\frac72}+14z^{-7}x^{\frac72}$\\
\hline
$\ket{3}\ket{3}$ & $16x^3+32x^4+z^2(15x^3+32x^4)+z^{-2}(15x^3+32x^4)+$\\
 & $z^4(12x^3+32x^4)+z^{-4}(12x^3+32x^4)+z^6(7x^3+32x^4)+z^{-6}(7x^3+32x^4)+$\\
 & $16z^8x^4+16z^{-4}x^4$\\
\hline
$\ket{7/2}\ket{7/2}$ & $x^{\frac72}(20z+20z^{-1}+18z^3+18z^{-3}+14z^5+14z^{-5}+8z^7+8z^{-7})$\\
\hline
$\ket{4}\ket{4}$ & $x^4(25+24z^2+24z^{-2}+21z^{4}+21z^{-4}+16z^6+16z^{-6}+9z^8+9z^{-8})$\\
\hline
\caption{$(SU(2)_1\times SU(2)_{-1})/\ZZ_2$}
\end{longtable}

\begin{longtable}{|l|l||l|l|}

  \hline
GNO charges & Index contribution & GNO charges & Index contribution\\
  \hline 
   $T=0$ & & $T=5$ & \\
 \hline
 $\ket{0,0}\ket{0,0}$ & $1+4x+12x^2+8x^3+12x^4$ & $\ket{3,2}\ket{3,2}$ & $2(6x^{\frac52}+14x^{\frac72})$ \\
 $\ket{1,-1}\ket{1,-1}$ & $4x+16x^2+16x^3+33x^4$ & $\ket{4,1}\ket{4,1}$ & $2(5x^{\frac52}+14x^{\frac72})$\\
 $\ket{1,-1}\ket{0,0}$ & $-x^4$ & $\ket{5,0}\ket{5,0}$ & $2(3x^{\frac52}+14x^{\frac7/2})$\\
 $\ket{0,0}\ket{1,-1}$ & $-x^4$ & $\ket{6,-1}\ket{6,-1}$ & $14x^{\frac72}$\\
 $\ket{2,-2}\ket{2,-2}$ & $9x^2+24x^3+16x^4$ & &\\
 $\ket{3,-3}\ket{3,-3}$ & $16x^3+32x^4$ & &\\
 $\ket{4,-4}\ket{4,-4}$ & $25x^4$ & &\\  
  \hline 
  $T=1$ & & $T=6$ & \\ 
 \hline
 $\ket{1,0}\ket{1,0}$ & $2(x^{\frac12}+6x^{\frac32}+10x^{\frac52}+7x^{\frac72})$ & $\ket{3,3}\ket{3,3}$ & $10x^3+16x^4$\\
 $\ket{2,-1}\ket{2,-1}$ & $2(3x^{\frac32}+10x^{\frac52}+8x^{\frac72})$ & $\ket{4,2}\ket{4,2}$ & $15x^3+32x^4$\\
 $\ket{3,-2}\ket{3,-2}$ & $2(6x^{\frac52}+14x^{\frac72})$ & $\ket{5,1}\ket{5,1}$ & $12x^3+32x^4$\\
 $\ket{4,-3}\ket{4,-3}$ & $20x^{\frac72}$ & $\ket{6,0}\ket{6,0}$ & $7x^3+32x^4$\\
 & &  $\ket{7,-1}\ket{7,-1}$ & $16x^4$\\  
  \hline
$T=2$ & & $T=7$ &\\
\hline
$\ket{1,1}\ket{1,1}$ & $3x+8x^2+12x^3+8x^4$ & $\ket{4,3}\ket{4,3}$ & $20x^{\frac72}$\\
$\ket{2,0}\ket{2,0}$ & $3x+16x^2+19x^3+24x^4$ & $\ket{5,2}\ket{5,2}$ & $18x^{\frac72}$\\
$\ket{3,-1}\ket{3,-1}$ & $8x^2+24x^3+16x^4$ & $\ket{6,1}\ket{6,1}$ & $14x^{\frac72}$\\
$\ket{4,-2}\ket{4,-2}$ & $15x^3+32x^4$ & $\ket{7,0}\ket{7,0}$ & $8x^{\frac72}$\\
$\ket{5,-3}\ket{5,-3}$ & $24x^4$ & &\\
\hline
$T=3$ & & $T=8$ & \\
\hline
$\ket{2,1}\ket{2,1}$ & $2(3x^{\frac32}+10x^{\frac52}+9x^{\frac72})$ & $\ket{4,4}\ket{4,4}$ & $15x^4$\\
$\ket{3,0}\ket{3,0}$ & $2(2x^{\frac32}+10x^{\frac52}+10x^{\frac72})$ & $\ket{5,3}\ket{5,3}$ & $24x^4$\\
$\ket{4,-1}\ket{4,-1}$ & $2(5x^{\frac52}+14x^{\frac72})$ & $\ket{6,2}\ket{6,2}$ & $21x^4$\\
$\ket{5,-2}\ket{5,-2}$ & $18x^{\frac72}$ & $\ket{7,1}\ket{7,1}$ & $16x^4$\\
 & & $\ket{8,0}\ket{8,0}$ & $9x^4$\\
\hline
$T=4$ & \\
\hline
$\ket{2,2}\ket{2,2}$ & $6x^2+12x^3+12x^4$ & &\\
$\ket{3,1}\ket{3,1}$ & $8x^2+24x^3+16x^4$ & &\\
$\ket{4,0}\ket{4,0}$ & $5x^2+24x^3+21x^4$ & &\\
$\ket{5,-1}\ket{5,-1}$ & $12x^3+32x^4$ & &\\
$\ket{6,-2}\ket{6,-2}$ & $21x^4$ & &\\
\hline
\caption{$U(2)_1\times U(2)_{-1}$.
$T$ stands for the topological charge.}
\end{longtable}

\begin{longtable}{|l|l|}
  \hline
GNO charges & Index contribution\\
  \hline 
$\ket{0}\ket{0}$ & $1+4x+12x^2+z^2(3x+8x^2)+z^{-2}(3x+8x^2)$\\
$\ket{1}\ket{1}$ & $4x+16x^2+z^2(3x+16x^2)+z^{-2}(3x+16x^2)$\\ 
$\ket{2}\ket{2}$ & $9x^2+8z^2x^2+8z^{-2}x^2$\\ 
\hline 
\caption{$SU(2)_1\times SU(2)_{-1}$}
\end{longtable}

\begin{longtable}{|l|l|}
  \hline
GNO charges & Index contribution\\
  \hline 
$\ket{0}\ket{0}$ & $1+4x+12x^2+z^2(3x+8x^2)+z^{-2}(3x+8x^2)$\\
$\ket{1/2}\ket{1/2}$ & $4x+17x^2+z^2(3x+16x^2)+z^{-2}(3x+16x^2)$\\ 
$\ket{1}\ket{1}$ & $9x^2+8z^2x^2+8z^{-2}x^2$\\ 
\hline 
\caption{$(SU(2)_2\times SU(2)_{-2})/\ZZ_2$}
\end{longtable}

\end{document}